# Generation and Transmission Expansion Planning with BESS-Based Virtual Transmission Lines

Qiushi Wang and Xingpeng Li
Department of Electrical and Computer Engineering
University of Houston
Houston, Texas, USA

***Abstract*—This paper proposes a mathematical model for long-term Generation and Transmission Expansion Planning (GTEP) that integrates a relaxed Virtual Transmission Line (VTL) as a Storage in Place of Transmission Asset (SIPTA) strategy to address challenges posed by transmission capacity shortages and system congestion in deregulated power markets, particularly under the rapid growth of renewable energy resources and integrated data centers. The proposed VTL formulation coordinates the operation of the two battery energy storage systems (BESSs) forming a VTL pair by preventing them from charging or discharging simultaneously, thereby providing additional congestion relief without introducing an explicit congestion cost into the objective function. Base case studies on a revised IEEE 24-bus system demonstrate that VTL reduces system congestion compared with the standalone BESS case, achieving a 12.3% reduction in weighted total congestion and up to approximately 67% reduction in congestion on individual transmission branches, with only a 0.3% increase in total planning cost. Furthermore, improvements across multiple congestion metrics indicate that the benefits of VTL extend beyond reducing congested line-hours to reducing congestion rent and N-1 transmission violations in the base case. These results demonstrate the potential of VTL to provide congestion mitigation that complements standalone BESS application.**

## Nomenclature

Set

| | |
|---|---|
| $\boldsymbol{D}$ | Set of days in the planning year. |
| $\boldsymbol{G}$ | Generators. |
| $\boldsymbol{G(n)}$ | Generators at bus *n*. |
| $\boldsymbol{J}$ | Candidate for new branches. |
| $\boldsymbol{J(n-)}$ | Candidate new branches of which bus *n* is from-bus. |
| $\boldsymbol{J(n+)}$ | Candidate new branches of which bus *n* is the to-bus. |
| $\boldsymbol{K}$ | Branches. |
| $\boldsymbol{K(n-)}$ | Branches of which bus *n* is the from-bus. |
| $\boldsymbol{K(n+)}$ | Branches of which bus *n* is the to-bus. |
| $\boldsymbol{N}$ | Buses. |
| $\boldsymbol{NGG}$ | New gas generators. |
| $\boldsymbol{NGG(n)}$ | New gas generators at bus n. |
| $\boldsymbol{NSG}$ | New solar generators. |
| $\boldsymbol{NSG(n)}$ | New solar generators at bus n. |
| $\boldsymbol{NWG}$ | New wind generators. |
| $\boldsymbol{NWG(n)}$ | New wind generators at bus n. |
| $\boldsymbol{NBESS}$ | New BESS. |
| $\boldsymbol{NBESS(n)}$ | New BESS at bus n. |
| $\boldsymbol{S}$ | Solar generations |
| $\boldsymbol{S(n)}$ | Solar generations at bus n. |
| $\boldsymbol{T}$ | Hours in a day. |
| $\boldsymbol{VT}$ | New VTL. |
| $\boldsymbol{W}$ | Wind generations |
| $\boldsymbol{W(n)}$ | Wind generations at bus n. |
| $\boldsymbol{Y}$ | A set of years in the planning horizon. |

Indices

| | |
|---|---|
| $\boldsymbol{d}$ | Day d, an element of set D. |
| $\boldsymbol{g}$ | Generator $g$, an element of set $G$. |
| $\boldsymbol{j}$ | Candidate new branch $j$, an element of set $J$. |
| $\boldsymbol{k}$ | Existing branch $k$, an element of set $K$. |
| $\boldsymbol{n}$ | Bus *n*, an element of the set $N$. |
| $\boldsymbol{n_{(k-)}}$ | From-bus of existing branch *k*. |
| $\boldsymbol{n_{(k+)}}$ | To-bus of existing branch *k*. |
| $\boldsymbol{n_{(j-)}}$ | From-bus of candidate new branch $j$. |
| $\boldsymbol{n_{(j+)}}$ | To-bus of candidate new branch $j$. |
| $\boldsymbol{e}$ | New battery *e*, an element of set $NBESS$ |
| $\boldsymbol{ngg}$ | New gas generator *ngg*, an element of set $NGG$ |
| $\boldsymbol{nsg}$ | New solar generator *nsg*, an element of set NSG |
| $\boldsymbol{nwg}$ | New wind generator *nwg*, an element of set NWG |
| $\boldsymbol{s}$ | Existing solar generator *s*, an element of set S |
| $\boldsymbol{t}$ | Hour $t$, an element of set $T$. |
| $\boldsymbol{v}$ | New virtual transmission *vt*, an element of set $VT$ |
| $\boldsymbol{w}$ | Existing wind generator *w*, an element of set W |
| $\boldsymbol{y}$ | Year *y*, an element of set $Y$. |

Parameters

| | |
|---|---|
| $\boldsymbol{\Delta T}$ | Time interval |
| $\boldsymbol{nT}$ | Hours of a day |
| $\boldsymbol{\alpha}$ | Annual load increase in percentage. |
| $\boldsymbol{\mu}$ | Annual maintenance cost in percentage for new branches. |
| $\overline{\boldsymbol{A_k^{ex}}}$ | Long-term limit of existing branch *k*. |
| $\overline{\boldsymbol{A_j^{new}}}$ | Long-term limit of candidate new branch $j$. |
| $\boldsymbol{C_g^{OP}}$ | Operational cost for unit $g$, \$. |
| $\boldsymbol{C_j^{CAP}}$ | Capital cost for constructing new branch $j$. |
| $\boldsymbol{M}$ | A very big constant number. |
| $\boldsymbol{N_y}$ | Number of years in a planning epoch. |
| $\overline{\boldsymbol{P_g}}$ | Maximum output of gas unit *g*. |
| $\underline{\boldsymbol{P_g}}$ | Minimum output of gas unit *g*. |
| $\overline{\boldsymbol{P_{ngg}^{inv}}}$ | Maximum available power to be invested for new gas unit *ngg*. |
| $\underline{\boldsymbol{P_{ngg}^{inv}}}$ | Minimum available power to be invested for new gas unit *ngg*. |
| $\underline{\boldsymbol{P_{ngg}}}$ | Minimum output of new gas unit *ngg*. |
| $\overline{\boldsymbol{P_{nsg}^{inv}}}$ | Maximum available power to be invested for new solar unit *nsg*. |
| $\underline{\boldsymbol{P_{nsg}^{inv}}}$ | Minimum available power to be invested for new solar unit *nsg*. |
| $\overline{\boldsymbol{P_{nwg}^{inv}}}$ | Maximum available power to be invested for new wind unit *nwg*. |
| $\underline{\boldsymbol{P_{nwg}^{inv}}}$ | Minimum available power to be invested for new wind unit *nwg*. |
| $\boldsymbol{D_{n,t,d}}$ | Load at bus *n* for hour $t$ on day d in the starting/first year. |
| $\boldsymbol{Y_n^{inv,IDC}}$ | Year in which the IDC load at bus *n* comes online. |
| $\boldsymbol{x_k^{ex}}$ | Reactance of existing branch *k*. |
| $\boldsymbol{x_j^{new}}$ | Reactance of candidate new branch $j$. |
| $\phi_{d,t}^{solar}$ | Solar power production capacity factor |
| $\phi_{d,t}^{wind}$ | Wind power production capacity factor |
| $C_j^{cap}$ | Capital cost for constructing new line j, $. |
| $C_e^{cap}$ | Capital cost for constructing new *battery* per 1 MWh energy capacity, $/MWh. |
| $C_g^{prod}$ | Variable production cost for gas unit *g* per 1 MWh, $/MWh. |
| $C_{ngg}^{cap}$ | Capital cost for constructing a new gas unit, g_gas, per 1 MW of power capacity, $/MW. |
| $C_{ngg}^{prod}$ | Variable production cost for new gas unit *ngg* per 1 MWh, $/MWh. |
| $C_{nsg}^{cap}$ | Capital cost for constructing new solar unit *g_solar* per 1 MW power capacity, $/MW. |
| $C_{nwg}^{cap}$ | Capital cost for constructing new wind unit *g_wind* per 1 MW power capacity, $/MW. |
| $P_{s,d,t}$ | Solar output for hour $t$ on day d of solar unit s. |
| $P_{w,d,t}$ | Wind output for hour $t$ on day d of wind unit w. |
| $D_{n,d,t,y}^{IDC}$ | IDC demand at bus n, hour t, day d, and year y. |
| $\overline{E_e}$ | Maximum available energy storage energy level for battery e. |
| $\underline{E_e}$ | Minimum energy storage energy level for battery e. |
| $T_e^c$ | Charging duration of energy storage battery e. |
| $T_e^d$ | Discharging duration of energy storage battery e. |
| $\overline{\sigma_e}$ | Maximum energy storage state of charge for battery e. |
| $\underline{\sigma_e}$ | Minimum energy storage state of charge for battery e. |
| $\eta_e^c$ | Charging efficiency for battery e. |
| $\eta_e^d$ | Discharging efficiency for battery e. |
| $\rho^{solar}$ | The solar curtailment penalty coefficient |
| $\rho^{wind}$ | The wind curtailment penalty coefficient |
| $L^{NL}$ | Expected lifespan of new transmission line |
| $L^{NGG}$ | Expected lifespan of new gas generator |
| $L^{NSG}$ | Expected lifespan of new solar |
| $L^{NWG}$ | Expected lifespan of new wind |
| $L^{NBESS}$ | Expected lifespan of new battery |

<u>*Variables*</u>:

| | |
|---|---|
| $P_{g,y,d,t}$ | Output power of gas unit *g* in year y, day d, hour t |
| $P_{ngg,y,d,t}$ | Gas generator power of new gas unit ngg in year y, day d, hour t |
| $P_{nsg,y,d,t}$ | Output power of new solar unit *nsg* in year y, day d, hour t |
| $P_{nwg,y,d,t}$ | Output power of new wind unit *nwg* in year y, day d, hour t |
| $P_{ngg}^{inv}$ | Gas generator power capacity should be invested. |
| $P_{nsg}^{inv}$ | Solar generator power capacity should be invested. |
| $P_{nwg}^{inv}$ | Wind generator power capacity should be invested. |
| $P_{s,y,d,t}^{curtail}$ | The curtailed solar generation s in year y, day d, hour t |
| $P_{w,y,d,t}^{curtail}$ | The curtailed wind generation w in the period t for year y. |
| $P_{e,y,d,t}^{c}$ | Charging rate of new battery e in year y, day d, hour t. |
| $P_{e,y,d,t}^{d}$ | Discharging rate of new battery e in year y, day d, hour t. |
| $E_{e,y,d,t}$ | Energy storage energy level in year y, day d, hour t. |
| $E_e^{inv}$ | Battery energy capacity should be invested. |
| $\theta_{n,y,d,t}$ | Phase angle of bus *n* in year y, day d, hour t. |
| $P_{k,y,d,t}$ | Power flow on existing line *k* in year y, day d, hour t. |
| $P_{j,y,d,t}$ | Power flow on new line *j* in year y, day d, hour t. |
| $u_{j,y}$ | Binary variable indicating the availability of candidate new line *j* in year *y*. |
| $u_{ngg,y}$ | Binary variable indicating the availability of candidate new gas generator ngg in year *y*. |
| $u_{nsg,y}$ | Binary variable indicating the availability of candidate new solar generator nsg in year *y*. |
| $u_{nwg,y}$ | Binary variable indicating the availability of candidate new wind generator nsg in year *y*. |
| $u_{e,y}$ | Binary variable indicating the availability of candidate new battery e in year *y*. |
| $u_{e,y,d,t}^{c}$ | 1 indicates charging mode; otherwise, 0. |
| $u_{e,y,d,t}^{d}$ | 1 indicates discharging mode; otherwise, 0. |
| $u_{v,y}$ | Binary variable indicating the availability of candidate new VTL v in year *y*. |
| $v_j$ | Binary variable indicating the investment of candidate new line *j*. |
| $v_{ngg}$ | Binary variable indicating the investment of candidate new gas generation ngg. |
| $v_{nsg}$ | Binary variable indicating the investment of candidate new solar generation nsg. |
| $v_{nwg}$ | Binary variable indicating the investment of candidate new wind generation nwg. |
| $v_e$ | Binary variable indicating the investment of candidate new battery e. |
| $v_v$ | Binary variable indicating the activation of candidate new virtual transmission v. |
| $C^{G,cap}$ | Total generator investment cost in the planning horizon. |
| $C^{G,op}$ | Total generation operating cost in the planning horizon. |
| $C^{T}$ | Total transmission investment and maintenance cost in the planning horizon. |
| $C^{BESS,cap}$ | Total battery investment cost in the planning horizon. |
| $C^{curtail}$ | Solar and wind curtailment penalty |
| $C^{LS}$ | Load shedding penalty |
| $R$ | Remaining value of the new transmission line, new gas generator, new solar, new wind, new battery at the end of the planning year |
| $R_j^{Line}$ | Remaining value of the new transmission line at the end of the planning year |
| $R_{ngg}^{NGG}$ | Remaining value of the new gas generator at the end of the planning year |
| $R_{nsg}^{NSG}$ | Remaining value of the new solar at the end of the planning year |
| $R_{nwg}^{NWG}$ | Remaining value of the new wind at the end of the planning year |
| $R_e^{NBESS}$ | Remaining value of the new battery at the end of the planning year |

# I. Introduction

A traditional goal of Transmission expansion planning (TEP) is to mitigate power system congestion, with substantial research focusing on congestion-based indicators to improve the solution of TEP [1] [2]. Following electricity market deregulation, however, the primary objective of TEP shifted from purely reliability-driven requirements to cost-based investment optimization [3] [4]. In a deregulated environment, transmission lines promote competition among generators, thereby maximizing social welfare. Consequently, in many cost-minimization TEP formulations under deregulated markets, the justification for transmission investment is based on financial analysis aimed at minimizing overall capital and operational costs, without incorporating explicit congestion-reduction metrics as separate objective terms. A parallel transformation has occurred in Generation Expansion Planning (GEP). Deregulated markets promote low-cost generation, contributing to the rapid growth of renewable energy resources (RES). This high penetration of RES shifts system operations and, in turn, creates uncertainty in TEP's financial analysis.

To capture the mutual dependencies between generation and transmission behavior, integrated Generation and Transmission Expansion Planning (GTEP) models have been proposed. Reference [5] and [6] proposed a long-term GTEP framework that simultaneously evaluates physical transmission lines, solar plants, wind farms, energy storage units, and flexible combined cycle gas turbines. Reference [7] provided a comprehensive review of operational constraints in unit commitment, long-term generation expansion planning, and energy system optimization models for studying power systems with high renewable penetration and presents a trade-off between computational time and model accuracy.

In addition to the stress of rapid clean energy integration, power grids are simultaneously facing unprecedented, aggressive load growth forecasts that have brought network congestion back to the forefront of grid operations [8]. Given that GTEP relies on economic minimization, it remains highly ambiguous whether a cost-justified expansion plan will effectively alleviate network congestion in a deregulated marketplace. This operational vulnerability is exacerbated by the breakthrough of large language models and the resulting widespread adoption of AI applications, which drives the explosive deployment of Integrated Data Centers (IDCs) as foreseen in major US Regional Transmission Organizations (RTOs). Based on ERCOT's adjusted long-term load forecast summer peak forecast, data center and crypto loads together are projected to add approximately 32.7 GW to ERCOT's system peak between 2025 and 2031, representing roughly 22.5% of the projected 2031 system peak demand [9].

A few mathematical models for data center load profiles have been developed for use in the expansion planning problem for data center location and capacities, many of which are detailed to cover both hardware-centric and software-centric energy consumption characteristics [10] [11] [12] [13] [14]. Reference [15] developed an interval multi-object optimization formulation to determine the expansion of the distribution system with the collaboration of the data center's demand response in market participation. Despite these advancements in localized data center modeling, little effort has been made to evaluate the system-level impact of massive IDC clustering on GTEP and system congestion. Moreover, once new physical transmission lines are justified, the practical route design, site acquisition, and infrastructure construction can take a decade to complete – creating a multi-year infrastructure mismatch during rapid IDC integration.

To fill the gap between the lack of transmission line capacity and the slow pace of transmission line permitting and deployment, non-wire alternatives have been adopted in many deregulated markets. One non-wire alternative is a Battery Energy Storage System (BESS), which can help with congestion management when properly designed and operated. However, regulation and market barriers hindered the deployment of storage technologies and have been largely overlooked in energy modeling research [16]. For example, Under Article 54 of Directive (EU) 2019/944, transmission System Operators (TSOs) are generally prohibited from owning or operating storage assets in order to promote competition and prevent monopolistic practices. Germany's National Regulatory Authority was among the first to approve an exception for Storage-As-Transmission-Assets (SATA) in 2019, as the German transmission system has been suffering from system congestion. The exception allows TSOs to own or operate BESS integrated into their system if they are "used for the sole purpose of ensuring a secure and reliable operation of the transmission or distribution system". Similar policy changes occurred across multiple deregulated markets, including Australia and the United States [17].

One application of large-scale batteries as a SATA is Grid Booster (GB), which follows a reactive system operation strategy that activates GBs only during line outages to relieve temporarily overloaded lines [18] [19]. GB is also known as the System Integrity Protection System (SIPS) service in Australia [20]. Although GBs increase transmission loading limit by providing a security N-1 support, the BESSs used as GBs do not participate in daily energy scheduling, thereby limiting their potential economic utilization relative to BESSs that can operate across both normal and contingency conditions. When the BESS-based Virtual Transmission Line (VTL) proposed as a SATA project, BESSs are installed at both ends of a physical line to mimic the line flow [21]. However, having the BESSs on both sides of the line absorb and inject energy at the same rate is overly strict and can yield infeasible optimization results.

Another concept that allows the BESS to provide value in daily operations through energy market is Storage in Place of Transmission Asset (SIPTA), which strategically site and operate BESS to shift or reduce the needs for transmission upgrade and expansion. The BESSs as a SIPTA project are sited near the load or on the load side of the transmission line to provide peak management and congestion relief. This application can facilitate more efficient utilization of generation resources. However, in a meshed system where the resource side and demand side of transmission line is blurry, identifying the boundary of a load center becomes a challenging task.

In our prior work [22], we proposed a relaxed VTL strategy as SIPTA to be utilized in meshed transmission system and evaluated its performance and the economic benefits of short-term operation in a power market environment. The relaxed VTL strategy removes the energy absorption and injection rates and adds a constraint to prevent BESSs on both sides of the transmission line from charging or discharging simultaneously. The research results concluded that the relaxed VTL strategy is the only solution comparable to physical transmission lines that can provide satisfying congestion relief and operational cost reduction without significantly sacrificing computing time or load payment compared to standalone cost-driven battery solutions, network reconfiguration, and physical transmission line solutions. Reference [23] further evaluated the benefits of VTL when considering Renewable Energy Sources uncertainty with a scenario-based stochastic security-constrained unit commitment model considering VTL, referred to as SSCUC-VTL. The simulation results demonstrate that VTL can achieve both operational cost reduction and congestion relief in a power system with solar and wind generation. While prior works [22] [23] focused on short-term day-ahead security-constrained unit commitment and compared operational cost, it is important to evaluate the investment costs and strategic capacity schemes of VTLs within a long-term capacity expansion framework.

A few studies discussed computational models for transmission expansion planning, considering the impact of a booster as a SATA project for inter-area applications [24] [25]. Their results demonstrate that VTL solutions can achieve significant cost reduction compared with conventional physical transmission line expansion or additional generation. However, these approaches require information on tie-line power flow, which limits their ability to participate directly in power markets. To date, the application and performance of relaxed VTLs as SIPTA within long-term GTEP have not been investigated. This paper aims to bridge the gap for the GTEP with relaxed VTL as SIPTA project.

The main contributions of this paper are: 1) a mixed-integer GTEP formulation that co-optimizes generation, physical transmission, standalone BESS, and relaxed VTL-paired BESS investment; 2) a planning-compatible relaxed VTL constraint set that avoids simultaneous charging or discharging of paired BESS units without enforcing equal power exchange; 3) a cost and congestion evaluation of BESS-based VTL against standalone BESS and physical transmission expansion benchmarks; and 4) a revised IEEE 24-bus case study incorporating data center load growth that quantifies the impact of VTL on system congestion, BESS investment timing, and physical transmission investment decisions..

The rest of the paper is organized in the following structure. Section II describes the detailed formulation of the GTEP models. Section III presents case studies using the proposed VTL model and discusses test results among different GTEP options. Finally, Section IV concludes the study and suggests future work.

## II. Methodologies

The proposed framework minimizes the total cost associated with the long-term expansion and operation of generation, physical transmission lines, BESSs, and BESS-based VTL coordination over a multi-year planning horizon. GTEP-PT considers generation and physical transmission expansion without BESS or VTL candidates. GTEP-BESS extends the GTEP-PT framework by allowing standalone BESS investment and operation, while retaining the same generation and physical transmission options. GTEP-VTL applies additional mandatory coordination restrictions to predefined VTL pairs when BESSs at both endpoints are installed. Thus, the three configurations share the same underlying GTEP formulation, while progressively adding BESS and VTL capabilities and constraints.

All three configurations use the same planning horizon, system operating constraints, data center integration and load growth assumptions, representative-day profiles, and formulations for renewable curtailment, load shedding, and residual asset value. The differences among the configurations arise from the available expansion and coordination options summarized in Table 1. The mathematical formulation and the configuration-specific BESS and VTL constraints are presented in the following subsections.

Table 1 Model Summary

| Model | Generation expansion | Physical transmission expansion | BESS | VTL coordination |
|---|---|---|---|---|
| GTEP-PT | ✓ | ✓ | — | — |
| GTEP-BESS | ✓ | ✓ | ✓ | — |
| GTEP-VTL | ✓ | ✓ | ✓ | ✓ |

### A. Objective Function

One objective function that suits all three GTEP problems can be formulated as equation (1) to minimize the total expected generator operation cost and infrastructure investment cost over the planning horizon with consideration of residual value, renewable curtailment and load shedding.

The operating cost of gas generators is given by equation (2), which calculates the weighted operating cost of existing and new gas generators corresponding to the represented days. The representative days are selected by k-medoids, while the day weights are represented by $\omega_d$. Equation (3) calculates the capital cost for new gas, solar, and wind generation, with the operating costs of solar and wind neglected because they do not incur fuel costs. Equation (4) calculates the one-time capital cost in addition to the annual maintenance cost for new physical lines. Maintenance costs are neglected for generation resources.

The total BESS investment cost $C^{BESS,cap}$ is proportional to the installed energy capacity $E_e^{inv}$, and is zero when no BESS candidates are available, as shown in equation (5). GTEP-BESS allows BESS investment and operation. GTEP-VTL uses the same objective as GTEP-BESS and differs only through the additional VTL coordination constraints. No separate VTL investment cost is included because the proposed VTL is implemented through coordinated operation of selected BESS pairs.

Power system assets produce a remaining value between the end of the contracted life and the end of the operating life [26]. To avoid penalizing assets that are still economically valuable at the end of the planning horizon, the objective function subtracts a residual value $R$ in (1). The residual value $R$ aggregates the individual residual values of every candidate asset class, including new transmission lines, new gas generators, new solar resources, new wind resources, and BESSs, which is defined in (6).

Objective:

$$min\ C^{G,op} + C^{G,cap} + C^{T} + C^{BESS,cap} - R + C^{curtail} + C^{LS} \tag{1}$$

where:

$$C^{G,op} = \sum_{y}\sum_{d}\omega_d\sum_{t}\Big(\sum_{g}P_{g,y,d,t}C_g^{prod} + \sum_{ngg}P_{ngg,y,d,t}C_{ngg}^{prod}\Big)\ \Delta T \tag{2}$$

$$C^{G,cap} = \sum_{ngg}P_{ngg}^{inv}C_{ngg}^{cap} + \sum_{nsg}P_{nsg}^{inv}C_{nsg}^{cap} + \sum_{nwg}P_{nwg}^{inv}C_{nwg}^{cap} \tag{3}$$

$$C^{T} = \sum_{j}v_j^{NL}C_j^{cap} + \sum_{j}\sum_{y}u_{j,y}^{NL}C_j^{cap}\mu \tag{4}$$

$$C^{BESS,cap} = \sum_{e\in NBESS}E_e^{inv}C_e^{cap} \tag{5}$$

$$R = \sum_{j\in J}R_j^{Line} + \sum_{ngg\in NGG}R_{ngg}^{NGG} + \sum_{nsg\in NSG}R_{nsg}^{NSG} + \sum_{nwg\in NWG}R_{nwg}^{NWG} + \sum_{e\in NBESS}R_e^{NBESS} \tag{6}$$

To encourage renewable energy utilization, renewable energy curtailment is incorporated into the objective function through the penalty term $C^{curtail}$. Curtailment of solar and wind generation is penalized at finite rates represented by the curtailment coefficient $\rho^{solar}$ and $\rho^{\mathrm{wind}}$, respectively, as defined in equation (7). The formulation distinguishes between existing and newly invested resources. To ensure a feasible economic solution, load shedding options are added to the GTEP formulations. Equation (8) is the load penalty.

$$C^{curtail} = \sum_y \sum_d \omega_d \sum_t (\rho^{solar} \sum_s P^{curtail}_{s,y,d,t} + \rho^{wind} \sum_w P^{curtail}_{w,y,d,t} + \rho^{solar} \sum_{nsg} P^{curtail}_{nsg,y,d,t} + \rho^{wind} \sum_{nwg} P^{curtail}_{nwg,y,d,t})\ \Delta T \quad (7)$$

$$C^{LS} = \rho^{load} \sum_{y,d,t,n} \omega_d D^{LS}_{n,t,d,y}\ \Delta T \quad (8)$$

### B. Generation and Transmission Expansion Planning

The objective function is subject to the power balance, line flow, and generator operating constraints. Equation (9) enforces the nodal power balance equation at each bus n, hour t, representative day d, and planning year y. Equation (10) defines the operating limits of existing conventional generators. The power flow on existing and candidate lines is modeled using the DC power flow equations (11) and (13), respectively, based on the B-θ formulation. The voltage angle at the designated reference bus is fixed at zero in all planning years and operating periods to provide the reference for the DC power flow formulation. Equations (12) and (14) enforce the thermal limits of the existing and new transmission lines, respectively. For candidate transmission lines, the power flow is constrained to zero unless the corresponding line has been constructed. Finally, Equations (15) and (16) define the transmission investment status, ensuring that once a candidate transmission line is commissioned, it remains in service for the remainder of the planning horizon.

$$\sum_{k\in K(n+)} P_{k,y,d,t} - \sum_{k\in K(n-)} P_{k,y,d,t} + \sum_{j\in J(n+)} P_{j,y,d,t} - \sum_{j\in J(n-)} P_{j,y,d,t} + \sum_{g\in G(n)} P_{g,y,d,t} + \sum_{ngg\in NGG(n)} P_{ngg,y,d,t} + \sum_{s\in S(n)} \left(P_{s,d,t} - P^{curtail}_{s,y,d,t}\right) + \sum_{w\in W(n)} \left(P_{w,d,t} - P^{curtail}_{w,y,d,t}\right) + \sum_{nsg\in NSG(n)} P_{nsg,y,d,t} + \sum_{nwg\in NWG(n)} P_{nwg,y,d,t} + \sum_{e\in NBESS(n)} P^{d}_{e,y,d,t} = (D^{LOAD}_{n,t,d,y} - D^{LS}_{n,t,d,y}) + \sum_{e\in NBESS(n)} P^{c}_{e,y,d,t},\ \ \forall n,t,d,y \quad (9)$$

$$\underline{P_g} \le P_{g,y,d,t} \le \overline{P_g} \quad \forall g,t,d,y \quad (10)$$

$$P_{k,y,d,t} = \frac{\theta_{n(k-),y,d,t} - \theta_{n(k+),y,d,t}}{x^{ex}_k} \quad \forall k,y,d,t \quad (11)$$

$$-\overline{A^{ex}_k} \le P_{k,y,d,t} \le \overline{A^{ex}_k} \quad \forall k,y,d,t \quad (12)$$

$$-M\left(1-u^{NL}_{j,y}\right) \le P_{j,y,d,t} - \frac{\theta_{n(j-),y,d,t} - \theta_{n(j+),y,d,t}}{x_j} \le M\left(1-u^{NL}_{j,y}\right)\ \forall j,y,d,t \quad (13)$$

$$-\overline{A^{new}_j} u^{NL}_{j,y} \le P_{j,y,d,t} \le \overline{A^{new}_j} u^{NL}_{j,y}\ \ \forall j,y,d,t \quad (14)$$

$$u^{NL}_{j,y-1} \le u^{NL}_{j,y} \quad \forall j,y>1 \quad (15)$$

$$v^{NL}_j \ge u^{NL}_{j,y} \quad \forall j,y>1 \quad (16)$$

$$v^{NL}_j \le \sum_y u^{NL}_{j,y} \quad \forall j,y \quad (17)$$

Equations (18)-(22) define constraints on the invested capacity and hourly distribution of new gas generators. It is assumed that once built in year y, the selected gas generator stays in service for the remaining years of the planning horizon, with its output limited to the invested capacity.

$$\underline{P_{ngg}} u_{ngg,y} \le P_{ngg,y,d,t} \le P^{inv}_{ngg} u_{ngg,y} \quad \forall ngg,t,d,y \quad (18)$$

$$\underline{P^{inv}_{ngg}} v_{ngg} \le P^{inv}_{ngg} \le \overline{P^{inv}_{ngg}} v_{ngg}\ \forall ngg \quad (19)$$

$$u_{ngg,y-1} \le u_{ngg,y} \quad \forall ngg,y>1 \quad (20)$$

$$v_{ngg} \ge u_{ngg,y} \quad \forall ngg,y>1 \quad (21)$$

$$v_{ngg} \le \sum_y u_{ngg,y} \quad \forall ngg,y \quad (22)$$

Equations (23)-(28) define constraints on new solar generator invested capacity. It is assumed that the sum of power delivered and curtailed by the new solar generation is proportional to the invested capacity and the solar power production capacity factor $\phi^{solar}$. The new solar curtailment constraints are given by equation (25).

$$P^{curtail}_{nsg,y,d,t} + P_{nsg,y,d,t} = P^{inv}_{nsg} \phi^{solar}_{d,t} u_{nsg,y}, \forall nsg,t,d,y \quad (23)$$

$$\underline{P^{inv}_{nsg}} v_{nsg} \le P^{inv}_{nsg} \le \overline{P^{inv}_{nsg}} v_{nsg}\ \forall nsg \quad (24)$$

$$0 \le P^{curtail}_{nsg,y,d,t} \le P^{inv}_{nsg} \phi^{solar}_{d,t} u_{nsg,y} \quad (25)$$

$$u_{nsg,y-1} \le u_{nsg,y} \quad \forall nsg,y>1 \quad (26)$$

$$v_{nsg} \ge u_{nsg,y} \quad \forall nsg,y>1 \quad (27)$$

$$v_{nsg} \le \sum_y u_{nsg,y} \quad \forall nsg,y \quad (28)$$

Equations (29)-(34) define constraints on new wind generator invested capacity. It is assumed that the sum of power delivered and curtailed by the wind generation is proportional to the invested capacity and the wind power production capacity factor $\phi^{wind}$. The new wind curtailment constraints are given by equation (31).

$$P^{curtail}_{nwg,y,d,t} + P_{nwg,y,d,t} = P^{inv}_{nwg} \phi^{wind}_{d,t} u_{nwg,y}, \forall nwg,t,d,y \quad (29)$$

$$\underline{P^{inv}_{nwg}} v_{nwg} \le P^{inv}_{nwg} \le \overline{P^{inv}_{nwg}} v_{nwg}\ \forall nwg \quad (30)$$

$$0 \le P^{curtail}_{nwg,y,d,t} \le P^{inv}_{nwg} \phi^{wind}_{d,t} u_{nwg,y} \quad (31)$$

$$u_{nwg,y-1} \le u_{nwg,y} \quad \forall nwg,y>1 \quad (32)$$

$$v_{nwg} \ge u_{nwg,y} \quad \forall nwg,y>1 \quad (33)$$

$$v_{nwg} \le \sum_y u_{nwg,y} \quad \forall nwg,y \quad (34)$$

The existing solar and wind generation curtailment constraints are defined by (35) and (36), respectively.

$$0 \le P_{s,y,d,t}^{curtail} \le P_{s,d,t}, \forall s,t,d,y \tag{35}$$

$$0 \le P_{w,y,d,t}^{curtail} \le P_{w,d,t}, \forall w,t,d,y \tag{36}$$

The power demand comprises the annual growth of the base load and newly built IDC loads as in equation (37). The IDC power consumption $D_{n,t,d,y}^{IDC}$ is phased in over the planning horizon and is modeled as a linear function of IT capacity utilization, accounting for both fixed overhead and dynamic computing workload as formulated in equation (38). $P_{n,y}^{IDC-idle}$ denotes the static baseline power drawn when the servers are idle. $P_{n,y}^{IDC-busy}$ reprepresents the full-load power consumption at maximum capacity. $\alpha_{t,d}^{IDC}$ is the time-varying workload factor representing the projected data center utilization profiles. The load shedding is bound by bus load in equation (39).

$$D_{n,t,d,y}^{LOAD} = D_{n,t,d}(1+\alpha)^{y-1} + D_{n,t,d,y}^{IDC} \tag{37}$$

$$D_{n,t,d,y}^{IDC} = \begin{cases} 0, & y < Y_n^{inv,IDC} \\ P_{n,y}^{IDC-idle} + (P_{n,y}^{IDC-busy} - P_{n,y}^{IDC-idle})\alpha_{t,d}^{IDC}, & y \ge Y_n^{inv,IDC}, \end{cases} \quad \forall n,t,d,y \tag{38}$$

$$0 \le D_{n,t,d,y}^{LS} \le D_{n,t,d,y}^{LOAD}, \forall n,t,d,y \tag{39}$$

Each investment's residual value is assumed to be in proportion to the useful life remaining at the planning horizon's end and is bounded by constraint equations (40)-(47) respectively.

$$R_j^{Line} \le (L^{NL} - \sum_y u_{j,y}) C_j^{cap} / L^{NL} \tag{40}$$

$$0 \le R_j^{Line} \le v_j C_j^{cap} \tag{41}$$

$$R_{ngg}^{NGG} \le (L^{NGG} - \sum_y u_{ngg,y}) P_{ngg}^{inv} C_{ngg}^{cap} / L^{NGG} \tag{42}$$

$$0 \le R_{ngg}^{NGG} \le v_{ngg} P_{ngg}^{inv} C_{ngg}^{cap} \tag{43}$$

$$R_{nsg}^{NSG} \le (L^{NSG} - \sum_y u_{nsg,y}) P_{nsg}^{inv} C_{nsg}^{cap} / L^{NSG} \tag{44}$$

$$0 \le R_{nsg}^{NSG} \le v_{nsg} P_{nsg}^{inv} C_{nsg}^{cap} \tag{45}$$

$$R_{nwg}^{NWG} \le (L^{NWG} - \sum_y u_{nwg,y}) P_{nwg}^{inv} C_{nwg}^{cap} / L^{NWG} \tag{46}$$

$$0 \le R_{nwg}^{NWG} \le v_{nwg} P_{nwg}^{inv} C_{nwg}^{cap} \tag{47}$$

### *C. BESS Model*

Consistent with how new transmission lines and new generators are treated for undepreciated value, BESS residual value is bounded by constraints in equation (48) and (49). The constraints assumed linearized capital cost depreciation.

$$R_e^{NBESS} \le (L^{NBESS} - \sum_y u_{e,y}) E_e^{inv} C_e^{cap} / L^{NBESS} \tag{48}$$

$$R_e^{NBESS} \le E_e^{inv} C_e^{cap} \tag{49}$$

In the nodal balance equation (9), each BESS unit injects its discharging power $P_{e,y,d,t}^{d}$ as a resource and withdraws its charging power $P_{e,y,d,t}^{c}$ as a load. BESS operation status is tied to investment status through equations (50)-(54). The charge and discharge mode indicator can be activated only if the BESS is installed on or before the operation year. It is assumed that once built in year y, the selected BESS stays in service for the following years in the planning horizon. Equation (55) indicates that BESS cannot charge and discharge simultaneously. Equations (56) and (57) cap the charging and discharging rates by invested capacity divided by charging and discharging duration. Equation (58) is the constraint for BESS investment capacity. Equation (59) bounds the state of energy, which is calculated by equation (60) with the cyclic boundary condition in equation (61). The state of energy is calculated based on the value from the previous time step.

$$u_{e,y}^{BESS} \ge u_{e,y,d,t}^{c} \quad \forall e,t,d,y \tag{50}$$

$$u_{e,y}^{BESS} \ge u_{e,y,d,t}^{d} \quad \forall e,t,d,y \tag{51}$$

$$u_{e,y}^{BESS} \ge u_{e,y-1} \quad \forall e, y > 1 \tag{52}$$

$$v_e^{BESS} \ge u_{e,y}^{BESS} \quad \forall e,y \tag{53}$$

$$v_e^{BESS} \le \sum_y u_{e,y}^{BESS} \quad \forall e,y \tag{54}$$

$$u_{e,y,d,t}^{c} + u_{e,y,d,t}^{d} \le 1 \quad \forall e,t,d,y \tag{55}$$

$$0 \le P_{e,y,d,t}^{c} \le \frac{E_e^{inv}}{T_e^c} u_{e,y,d,t}^{c}, \forall e,t,d,y \tag{56}$$

$$0 \le P_{e,y,d,t}^{d} \le \frac{E_e^{inv}}{T_e^d} u_{e,y,d,t}^{d}, \forall e,t,d,y \tag{57}$$

$$\underline{E_e} v_e^{BESS} \le E_e^{inv} \le \overline{E_e} v_e^{BESS}, \forall e \tag{58}$$

$$E_e^{inv} \underline{\sigma_e} u_{e,y}^{BESS} \le E_{e,y,d,t} \le E_e^{inv} \overline{\sigma_e} u_{e,y}^{BESS}, \forall e,t,d,y \tag{59}$$

$$E_{e,y,d,t} = E_{e,y,d,t-1} + (\eta_e^c P_{e,y,d,t}^c - P_{e,y,d,t}^d / \eta_e^d) \Delta T, \forall e,t,d,y \tag{60}$$

$$E_{e,y,d,0} = E_{e,y,d,nT}, \forall e,d,y \tag{61}$$

### *D. Virtual Transmission Model*

The virtual transmission line to be modeled in this section is the relaxed VTL as SIPTA project. The paired BESS candidates of VTLs are a subset of new BESS set, $NBESS_v \subseteq NBESS$, and governed by all BESS constraints as described in section II.C.

The paired BESS coordinate with each other on charging and discharging performance and can help reduce congestion risks of the existing transmission system. The two paired BESS of the VTL are not allowed to charge simultaneously when VTL is activated as in equation (62) and not allowed to discharge simultaneously when VTL is activated as in equation (63). Equations (64) ensure that a VTL is activated only when both paired BESSs are invested and operational in year y. Equation (65) forces the VTL mode once both BESSs in the pair are operational. The VTL activation status indicator is tied to its yearly operational status indicator by equation (66) and (67).

$$u_{e1,y,d,t}^{c} + u_{e2,y,d,t}^{c} \le 2 - u_{v,y}^{VTL}, \forall v,t,d,y \tag{62}$$

$$u_{e1,y,d,t}^{d} + u_{e2,y,d,t}^{d} \le 2 - u_{v,y}^{VTL}, \forall v,t,d,y \tag{63}$$

$$u_{v,y}^{VTL} \le u_{e,y}^{BESS}, e \in NBESS_v, NBESS_v \subseteq NBESS, \forall v,y \tag{64}$$

$$u_{v,y}^{VTL} \ge \sum_{e \in NBESS_v} u_{e,y}^{BESS} - 1, \qquad NBESS_v \subseteq NBESS, \forall v,y \tag{65}$$

$$v_v^{VTL} \geq u_{v,y}^{VTL} \ \forall v, y \tag{66}$$

$$v_v^{VTL} \leq \sum_y u_{v,y}^{VTL} \ \forall v, y \tag{67}$$

## III. Case Studies

The proposed relaxed VTL model for GTEP is tested and evaluated using a revised IEEE 24-bus system, as shown in Figure 1. The test tool is developed using the Gurobi solver in the Python-based Pyomo package. The nonlinear constraints in the equations can be solved by Gurobi directly as quadratic constraints. The Python script was run in a PyCharm environment with python 3.7 interpreter on Intel(R) Xeon(R) W-2195 CPU @ 2.30GHz system.

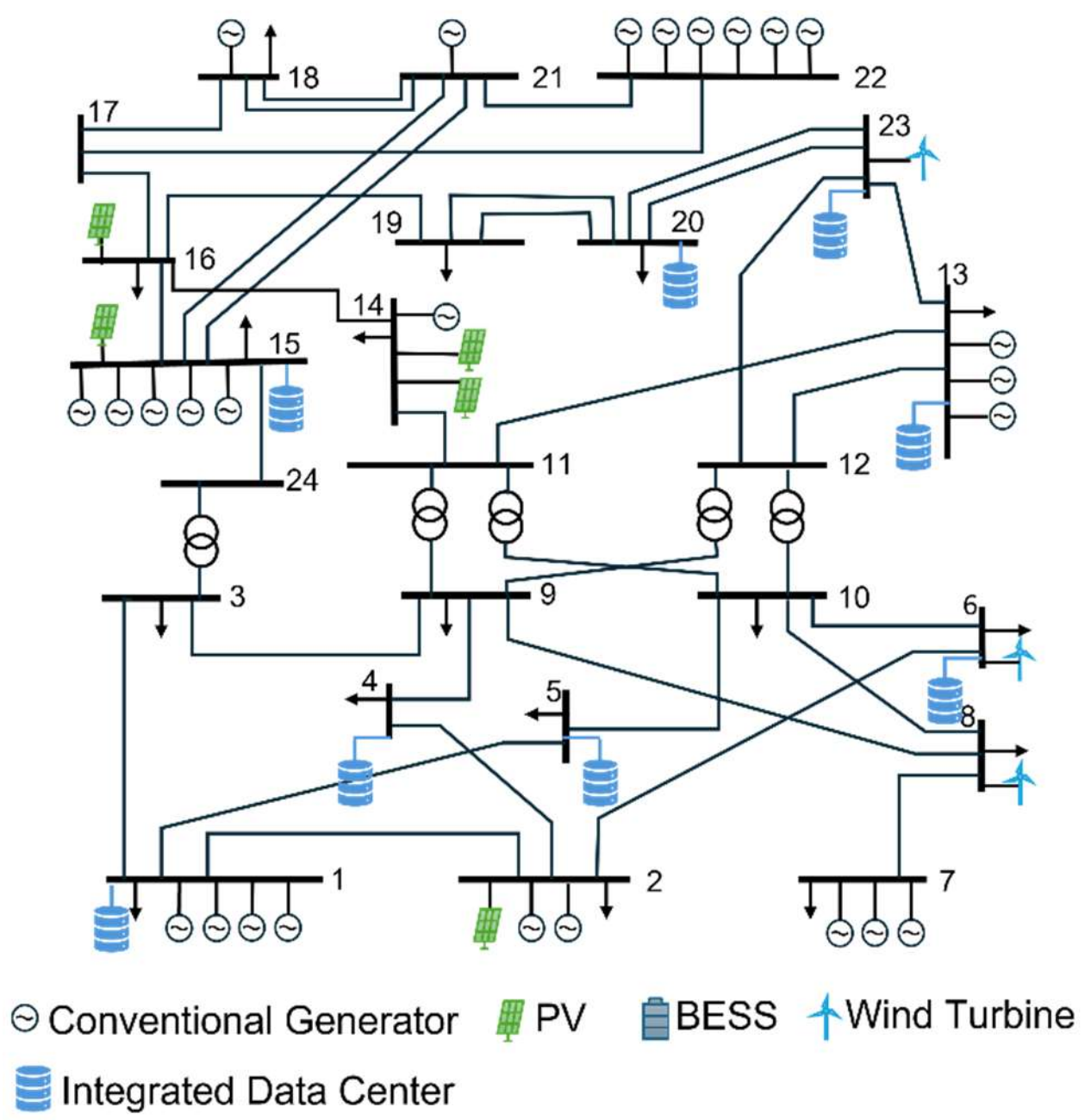

Figure 1 Revised IEEE 24-bus system.

### A. Test System Description

The revised IEEE 24-bus test system has an initial peak load of 5,101 MW and a total installed generation capacity of 8,482 MW. Conventional coal generation units at buses 2, 4, 14, 15, 16, and 23 are replaced with solar and wind generation at buses 2, 6, 8, 14, 15, 16, 23, increasing the share of existing renewable generation to approximately 43.6% of the total installed generation capacity. The minimum generation outputs of all conventional generators are set to zero in the GTEP case study. Thus, although the formulation includes generator minimum-output constraints, these constraints do not impose must-run generation in numerical experiments. Over the 10-year Generation and Transmission Expansion Planning (GTEP) horizon, a total of 2,600 MW of Integrated Data Center (IDC) loads is introduced at buses 1, 4, 5, 6, 13, 15, 20, and 23 along the 10 year horizon. Additionally, the remaining non-IDC system demand is assumed to grow at an annual rate of 4%. Under these assumptions, the total system peak load reaches 9,860 MW by the end of the planning horizon.

To construct a tractable yet representative time horizon for the multi-year GTEP model, we applied a k-medoids clustering procedure with forced extreme-day inclusion to the 366 daily load profiles of the 2020 RTS-GMLC dataset [27]. Each day was modeled as a 24-hour load vector. The annual peak-load day and the day with the largest hour-to-hour ramp were fixed as medoids, while remaining four medoids were selected using farthest-point initialization and greedy swap improvement. Cluster cardinalities were used as representative-day weights. The same K=6 medoid dates were applied to solar and wind profiles to preserve temporal consistency. Candidate investment options include physical transmission lines, gas turbines, solar plants, wind farms, and BESS. Capital costs for these technologies are derived from the surveys documented in [28] [29] [30].

### B. Study Results

Table 2 summarizes the optimal investment decisions obtained for the three test cases: GTEP-PT, GTEP-BESS and GTEP-VTL, using all available threads and with a relative optimality gap of 1%. The visualization of the cumulative investment capacity and investment timing of each applicable asset type are shown in Figure 2 and Figure 3, respectively.

Table 2 Optimal Investment Decisions for GTEP-PT, GTEP-BESS and GTEP-VTL Cases

| Asset | GTEP-PT | GTEP-BESS | GTEP-VTL |
|---|---|---|---|
| **New branch** | Y2: 18-19 | Y5: 18-19;<br>Y4: 16-17 | Y5: 18-19;<br>Y4: 16-17 |
| **New gas** | 2,485 MW | 1,202 MW | 1,444 MW |
| **New solar** | 3,965 MW | 5,136 MW | 4,934 MW |
| **New wind** | 0 MW | 0 MW | 0 MW |
| **BESS** | None | 6,635 MWh | 5,359 MWh |
| **VTL** | None | None | Y3: 10–14 |

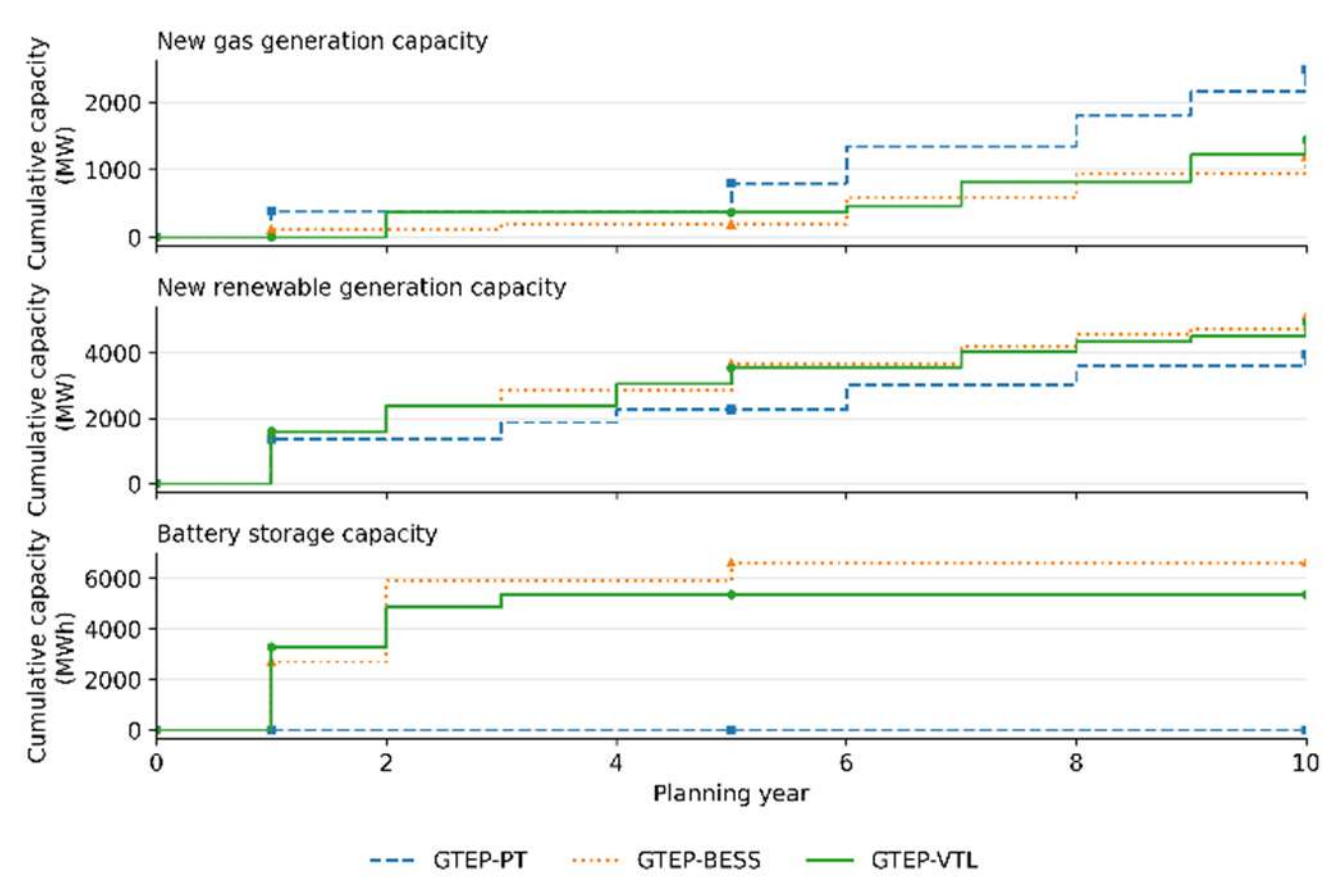

Figure 2 Accumulated new generation and BESS investment capacity through the planning horizon

Compared with GTEP-PT, both GTEP-BESS and GTEP-VTL require one additional new transmission line investment, but with a delayed deployment timeline. The introduction of batteries alters the optimal gas and solar investment portfolio: compared with GTEP-PT case, both GTEP-BESS and GTEP-VTL tend to shift from new gas generation toward new solar generation when possible. The GTEP-VTL case activates one pair of VTLs in Year 3. The VTL operation constraints result in an approximately 19% reduction in BESS investment capacity in GTEP-VTL compared with GTEP-BESS.

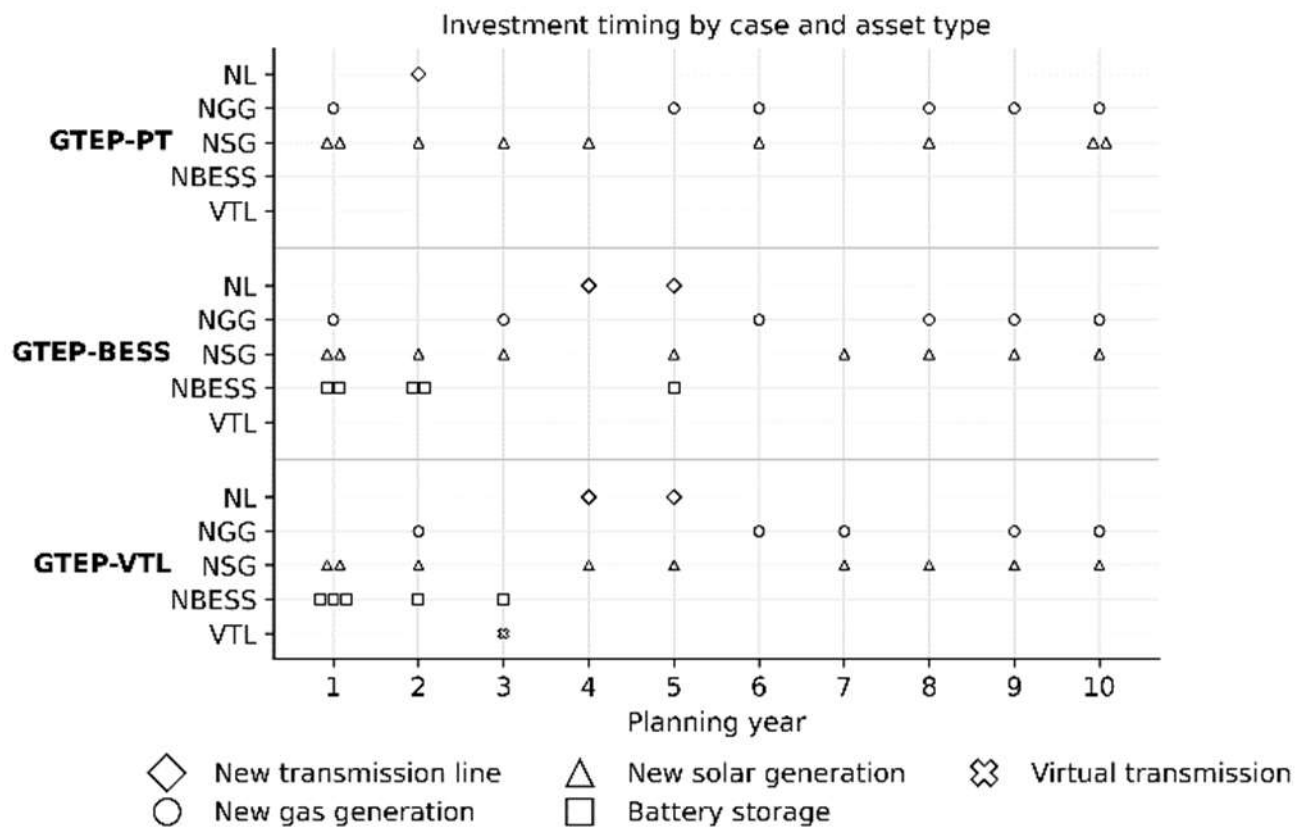

Figure 3 Investment decision and installation year for all asset types for GTEP-PT, GTEP-BESS, and GTEP-VTL cases

In addition, the VTL constraints alter the charging and discharging coordination of the BESSs, as illustrated in Figure 4. Under GTEP-VTL, BESSs at buses 10 and 14 do not charge simultaneously nor discharge simultaneously, whereas simultaneous charging or discharging occurs in the GTEP-BESS case. By preventing simultaneous charging or discharging at these locations, VTL changes the spatial distribution of BESS power injections and withdrawals and consequently modifies the power-flow patterns in the network. This provides a mechanism for reducing congestion without explicitly introducing additional congestion-related constraints or objectives into the optimization model.

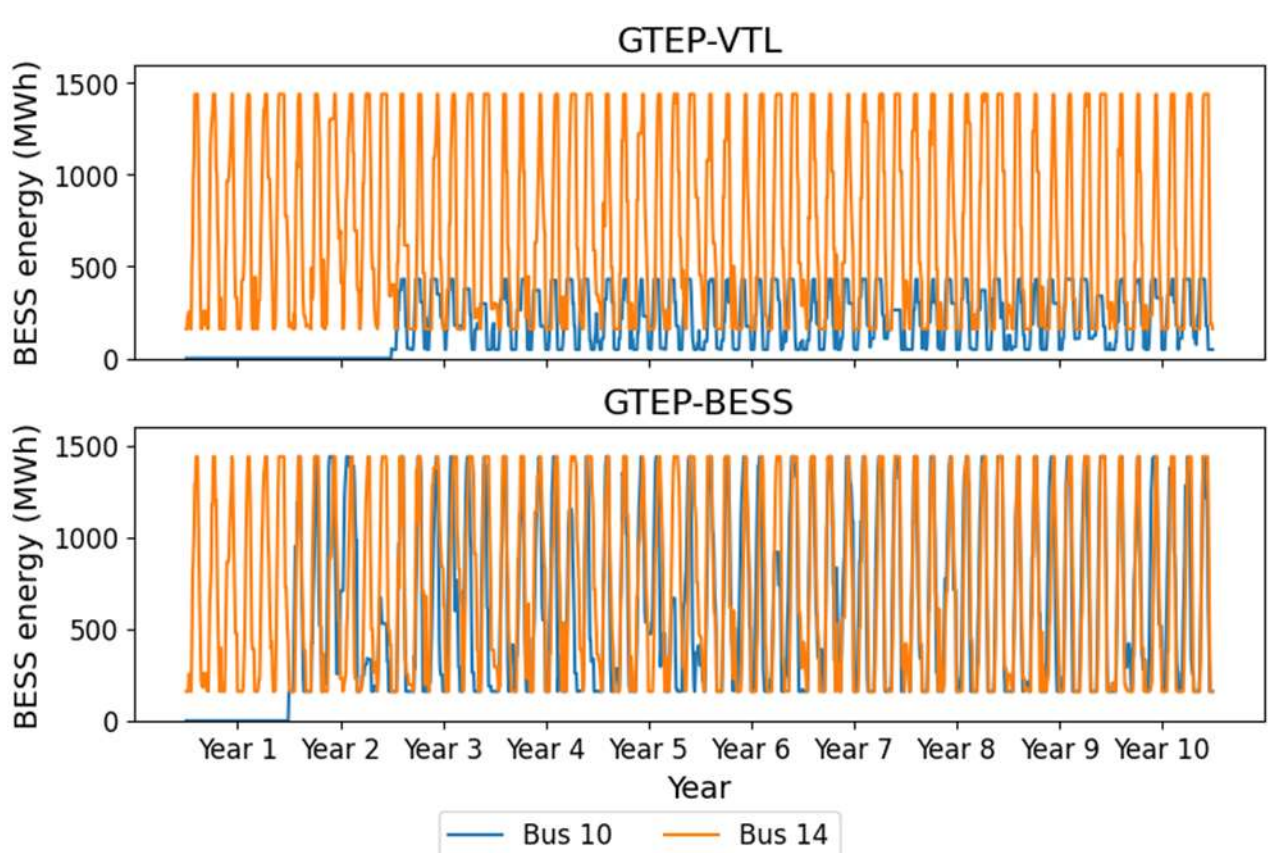

Figure 4 Stored-energy trajectories of the BESSs at Bus 10 and 14 under the GTEP-VTL and GTEP-BESS cases.

The overall performance of the three test cases, GTEP-PT, GTEP-BESS and GTEP-VTL, is summarized in Table 3. GTEP-BESS is used as the normalization baseline (100%) for comparing the GTEP-BESS and GTEP-VTL configurations. The cost details for the three cases are illustrated in Figure 5.

Table 3 Comparison of the cost and congestion for test cases

| | GTEP-PT | GTEP-BESS | GTEP-VTL |
|---|---|---|---|
| **Total Cost ($MM)** | 7,605 (103.3%) | 7,360 (100%) | 7,380 (100.3%) |
| **Renewable Curtailment Penalty ($MM)** | 36.4 | 4.4 | 9.7 |
| **Load Shedding Penalty ($MM)** | 3.41 | 1.03 | 2.74 |
| **Total $CO_2$ Emission (Mt)** | 48.97 | 43.75 | 44.47 |
| **System Total Congestion (weighted)** | 164,424 (70.4%) | 233,544 (100%) | 204,825 (87.7%) |
| **Total Congestion Rent ($MM)** | 1,265 (84.6%) | 1,495 (100%) | 1,453 (97.2%) |
| **N-1 Violation Instances (weighted)** | 900,721 (63.2%) | 1,425,516 (100%) | 1,303,543 (91.4%) |

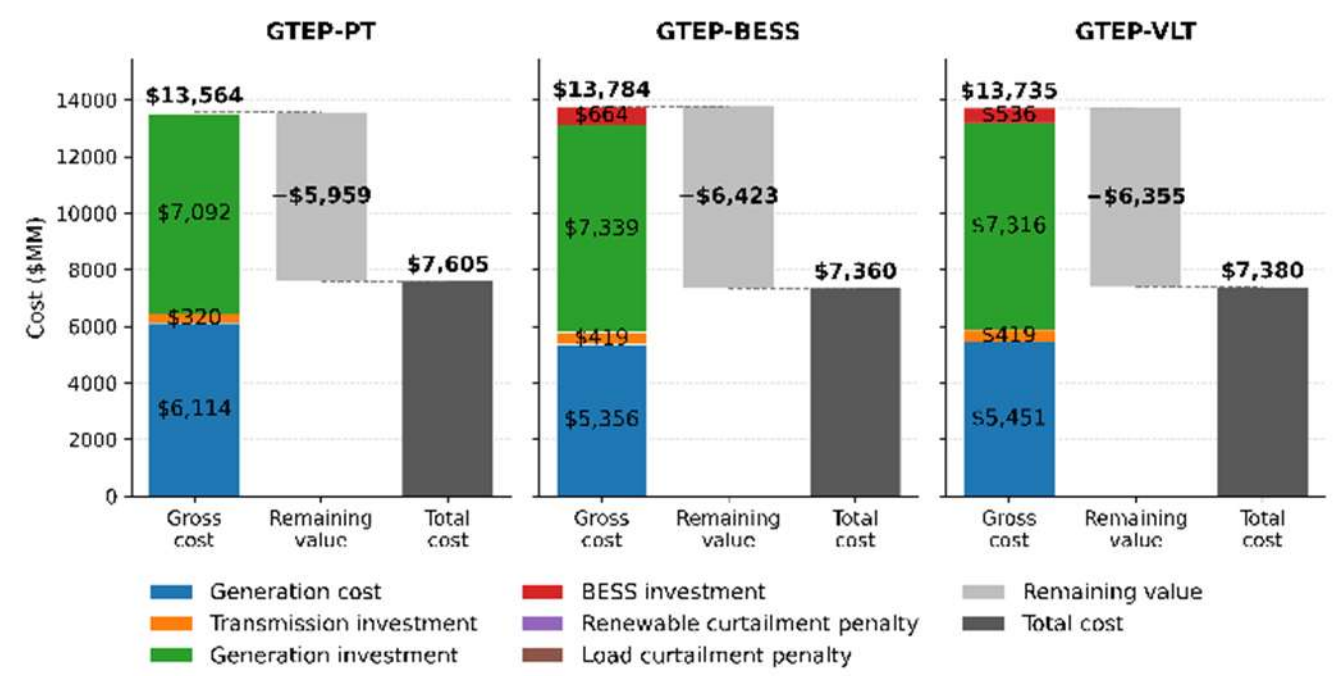

Figure 5 Cost Details and Comparison

The total cost of GTEP-BESS and GTEP-VTL are comparable and within the 1% optimization gap. GTEP-PT has 3% higher total cost than GTEP-BESS, primarily contributed by higher generation operation cost. With greater investment in gas generation and power delivery, GTEP-PT also has the highest total $CO_2$ emission. Adding batteries to the system can help promote renewable investment and reduce renewable curtailment and load shedding by providing additional temporal and locational flexibility, as demonstrated by GTEP-BESS and GTEP-VTL results.

In this analysis, system congestion is evaluated using three metrics: weighted system total congestion, total congestion rent, and weighted N-1 violation instances.

The weighted system total congestion represents the day-weighted line-hours in which a branch reaches its limit, summed over all branches across 24 hourly profiles over the 10-year planning horizon.

Congestion rent is computed from the dual variables of a linear programming problem obtained from the original solution by fixing all binary variables (investment decisions, hourly BESS charge/discharge mode decisions, and VTL activation decisions) and all continuous investment capacity decisions at their solution values. Under the lossless DC power flow model, total congestion rent is calculated as the thermal-limit dual multiplied by the line rating.

N-1 violation instances are identified through preventive DC screening based on Line Outage Distribution Factors (LODFs). The LODFs correspond to the network commissioned in each year. An N-1 violation instance is flagged when the line flow exceeds the line thermal, assuming no redispatch. The weighted violation instances are accumulated across 24 hourly profiles over the 10-year planning horizon for all branches in the system.

GTEP-BESS achieves the lowest total cost, while GTEP-PT exhibits the lowest system-wide congestion among the three cases. In comparison, GTEP-VTL achieves a substantial reduction in congestion relative to GTEP-BESS with only a 0.3% increase in total cost. Specifically, GTEP-VTL reduces weighted congested line-hours by 12.3%, total congestion rent by 2.8%, and weighted N-1 violation instances by 8.6%.

Table 4 presents the congestion reductions on individual transmission branches resulting from the introduction of VTL. Only one VTL pair in the optimized solution is activated, pairing BESSs on buses 10–14. The branches experiencing the largest congestion reductions do not necessarily connect directly to the VTL buses. This is because the deployment of VTL changes power-flow patterns throughout the surrounding network and can consequently relieve congestion on adjacent transmission branches. Four branches see congestion reduced by more than 14%, reaching as high as 67.76% on one branch, which demonstrates that VTL's congestion-relief effects are not confined to the branches directly linked to its deployment locations but extend across the surrounding network.

Table 4 Summary of Weighted Line Congestion for the Five Transmission Lines with the Largest Percentage Reduction under GTEP-VTL

| Branch (from bus – to bus) | GTEP-BESS weighted congestion | GTEP-VTL weighted congestion | Reduction (%) |
|---|---|---|---|
| **1-2** | 8,219 | 2,650 | 67.76% |
| **6-10** | 10,569 | 6,286 | 40.52% |
| **8-9** | 31,716 | 20,441 | 35.55% |
| **15-16** | 15,463 | 13,289 | 14.06% |
| **11-14** | 32,707 | 30,641 | 6.32% |

Table 5 presents the five most congested branches under the GTEP-BESS solution. The results show that the impact of VTL is not uniform across the network: congestion is substantially reduced on some of the most congested branches, while other highly congested branches experience little change or a slight increase in congestion. This heterogeneous response reflects the network-wide redistribution of power flows induced by VTL. Therefore, VTL locations should be evaluated through a system-wide optimization, rather than selecting VTL pairs based solely on the congestion level of individual branches, to capture the interactions among VTLs and the adjacent network.

Table 5 Summary of Weighted Line Congestion for The Top Five Congested Transmission Lines Under the GTEP-BESS and GTEP-VTL Solutions

| Branch (from bus – to bus) | GTEP-BESS weighted congestion | GTEP-VTL weighted congestion | Reduction (%) |
|---|---|---|---|
| **8-10** | 34,664 | 32,639 | 5.84% |
| **11-14** | 32,707 | 30,641 | 6.32% |
| **8-9** | 31,716 | 20,441 | 35.55% |
| **15-16** | 15,463 | 13,289 | 14.06% |
| **16-17** | 14,100 | 14,298 | -1.40% |

### *C. Sensitivity Analysis – BESS Capital Cost*

To evaluate the robustness and the limitations of the proposed GTEP-VTL scheme, sensitivity analysis was conducted across BESS capital costs. Table 6 compares system costs and congestion for GTEP-VTL and GTEP-BESS cases for BESS capital cost as 130% of the base value. VTL is activated and provides additional congestion relief at a cost of higher renewable curtailment penalty.

Table 6 Comparison of the cost and congestion for test cases with 130% BESS capital cost

| | GTEP-BESS | GTEP-VTL |
|---|---|---|
| **Total Cost ($MM)** | 7,460 (100%) | 7,468 (100.1%) |
| **Renewable Curtailment Penalty ($MM)** | 7.88 | 22.33 |
| **Load Shedding Penalty ($MM)** | 1.99 | 0.52 |
| **Total $CO_2$ Emission (Mt)** | 45.97 | 46.00 |
| **System Total Congestion (weighted)** | 183,038 (100%) | 177,068 (96.7%) |
| **Total Congestion Rent ($MM)** | 1,356 (100%) | 1,159 (85.5%) |
| **N-1 Violation Instances (weighted)** | 1,184,705 (100%) | 1,163,043 (98.2%) |

### *D. Sensitivity Analysis – Unavailability of new Pyhsical Transmission Line*

To evaluate the impact of unavailability of new physical transmission line, the new transmission line candidates are removed from the available asset options. Table 7 compares the cost and congestion for test cases without new physical line option. Without new transmission investment, GTEP-VTL achieves lower system-wide congestion and fewer N-1 violation instances than GTEP-BESS, but results in higher total congestion rent. The increase in congestion rent is primarily driven by storage investment timing. The VTL coordination constraint reduce the marginal value of BESS at bus 10 (one of the VTL pair node), resulting in approximately 60% less BESS capacity at bus 10 and delaying its deployment from Year 1 to Year 4; BESS at bus 14 is also delayed by one year to Year 2. Consequently, in Year 1, before storage is available at buses 10 and 14, the system-wide congestion rent in GTEP-VTL is $176 MM higher than in the GTEP-BESS case ($271 MM vs. $95 MM). Branch 6–10 alone accounts for $116 MM of the $176 MM excess, as all peak demand at bus 10 must be served through that branch without BESS buffering. Over Years 2-10, GTEP-VTL reduces congestion rent by $89 MM relative to GTEP-BESS.

Table 7 Comparison of the cost and congestion for test cases without new physical line option

| | GTEP-BESS | GTEP-VTL |
|---|---|---|
| **Total Cost ($MM)** | 7,441 (100%) | 7,449 (100.1%) |
| **Renewable Curtailment Penalty ($MM)** | 17.74 | 15.44 |
| **Load Shedding Penalty ($MM)** | 1.76 | 2.74 |
| **Total $CO_2$ Emission (Mt)** | 43.60 | 44.48 |
| **System Total Congestion (weighted)** | 213,185 (100%) | 200,614 (94.1%) |
| **Total Congestion Rent ($MM)** | 1,689 (100%) | 1,776 (105.2%) |
| **N-1 Violation Instances (weighted)** | 1,344,251 (100%) | 1,318,736 (98.1%) |

## IV. CONCLUSION

This research highlights the role of Virtual Transmission Lines (VTLs) in modern long-term power system planning, particularly when traditional physical transmission expansion is hindered by long permitting and construction timelines. This paper adopts a practical relaxed VTL as Storage in Place of Transmission Asset (SIPTA) strategy and develops models to incorporate it into a long-term generation and transmission expansion planning (GTEP) framework. Case studies on the revised IEEE 24-bus system demonstrate that VTL can provide complementary benefits to standalone BESS.

While both BESS and VTL help alleviate the need for near-term transmission line investment and reduce RES curtailment and load shedding, VTL achieves fewer congested line-hours and N-1 violation instances, with only a marginal change in total planning cost compared with standalone BESS. In both the base case and the BESS cost sensitivity study, GTEP-VTL also reduces congestion rent relative to GTEP-BESS. In the sensitivity study considering the unavailability of new physical transmission line, however, GTEP-VTL results in higher congestion rent in the early planning year due primarily to differences in the timing of BESS deployment. By altering the spatial coordination of BESS charging and discharging, VTL changes network power flow patterns and can alleviate congestion on neighboring transmission branches. These findings demonstrate that VTL can provide congestion mitigation that complements standalone BESS.

Future work will investigate VTL performance under larger and more diverse power-system test cases, as well as more detailed operational and investment assumptions.